\documentclass[twocolumn,prl,showpacs,amsmath,amssymb,floatfix]{revtex4-1}
\usepackage{graphicx,hyperref,bm}
\pdfoutput=1
\newcommand{\UU}{U}
\newcommand{\UUscat}{U_\text{s}}
\newcommand{\UUrand}{U_\text{r}}
\newcommand{\U}{\mbox{U}}
\newcommand{\EE}{\mathbb{E}}
\newcommand{\cb}{\bm{c}}
\newcommand{\rb}{\bm{r}}

\newcommand{\ZZ}{\mathcal{Z}}

\DeclareMathOperator{\STr}{STr}
\DeclareMathOperator{\Det}{Det}

\begin{document}
\title{Pure scaling operators at the integer quantum Hall plateau transition}
\author{R.~Bondesan, D.~Wieczorek, M.R.~Zirnbauer\\
  \textit{Institut f\"ur Theoretische Physik, Universit\"at zu K\"oln,
    Z\"ulpicher Stra{\ss}e 77, 50937 K\"oln, Germany}}
\begin{abstract}
Stationary wave functions at the transition between plateaus of the
integer quantum Hall effect are known to exhibit multi-fractal
statistics. Here we explore this critical behavior for the case of
scattering states of the Chalker-Coddington network model with point
contacts. We argue that moments formed from the wave amplitudes of
critical scattering states decay as pure powers of the distance
between the points of contact and observation. These moments in the
continuum limit are proposed to be correlation functions of primary
fields of an underlying conformal field theory. We check this proposal
numerically by finite-size scaling. We also verify the CFT prediction
for a 3-point function involving two primary fields.
\end{abstract}

\pacs{72.15.Rn, 73.43.-f, 73.43.Nq, 05.45.Df}

\maketitle

\paragraph{Introduction.} A revealing monitor of quantum critical
behavior driven by disorder is multi-fractal wave-function
statistics. In this vein, theory and experiment have focused on the
multi-fractality at Anderson localization transitions between
different topological phases of disordered electrons in two
dimensions, the prime example being the transition between plateaus of
the Hall conductance in the integer quantum Hall (IQH) effect
\cite{Evers2008}.

There has long been a consensus that it should be possible to describe
the IQH transition by a conformal-invariant effective field
theory. Yet, in spite of many efforts
\cite{Bhaseen2000,Ikhlef2011,Bettelheim2012} it remains an unsolved
problem to identify that conformal field theory (CFT) description. To
make progress with the search for it, one needs to find the conformal
fields and determine their scaling dimensions. A step in this
direction was taken in \cite{Janssen1999,Klesse2001}, where the
moments of the point-contact conductance were introduced and studied
as correlation functions. Alas, these are coherent sums of conformal
field correlators and therefore do not give direct access to
individual conformal fields in pure form; see \cite{Obuse2013} for a
recent discussion.

The purpose of this Letter is to put forth a large (and so far
unrecognized) class of multi-fractal observables that correspond
directly to correlators of CFT primary fields. Our results are
motivated by a recent $\sigma$-model based classification of scaling
fields at Anderson transitions \cite{Gruzberg2011,Gruzberg2013}. The
new feature here is that we focus on the scattering states of an
\emph{open} system, while the previous work concerned 
moments of the local density of states for \emph{closed} systems. For
concreteness and simplicity, we work with the Chalker-Coddington (CC)
network model.

The CC model is known to be related by a duality transformation to a
statistical mechanical system of vertex-model type
\cite{Gruzberg1997,Zirnbauer1997}. The main advance of our work is to
construct lattice approximations for pure scaling fields on both sides
of the duality -- as scattering observables of the CC model and,
equivalently, as operators of the vertex model. Both representations
serve a purpose. Based on the latter, we argue that our lattice
operators indeed are discretizations of pure scaling fields, while the
former makes it possible to compute their conformal dimensions
numerically by finite-size scaling.

\paragraph{CC model and scattering states.} We begin with a quick
review of the CC model \cite{Chalker1988}. This is a network model for
the quantum dynamics of an electron moving in two dimensions under the
influence of a strong magnetic field and a random electric
potential. Formulated on a square lattice, the model is built from
elementary plaquettes with a definite sense of circulation that
alternates between neighboring plaquettes. The links of the network
are directed accordingly, so that each site has two incoming and two
outgoing links. The electron wave function lives on the links and
evolves in discrete time as $\vert \psi(t+1) \rangle = \UU \vert
\psi(t) \rangle$ by a unitary operator $\UU = \UUscat\, \UUrand$. The
factor $\UUrand$ is a diagonal matrix modeling the propagation along
the links; it assigns to each link a random, independent and uniformly
distributed $\U(1)$ phase. The factor $\UUscat$ is non-random and
consists of $2 \times 2$ matrices that describe the transfer from
incoming to outgoing links at each site. When the probabilities for
transfer to the left/right are equal, the model is critical and
falls into the universality class of the IQH transition
\cite{Chalker1988}.

While it is of some interest to study the spectral properties and
stationary wave function statistics of the closed network, here we
turn to an open network. One major advantage of the open setting is
that it allows one to formulate and study CFT correlators right at the
critical point. (In contrast, Green's functions of the closed system
are defined by introducing a regularization which places the system
slightly off criticality.)

The network is opened up by severing a subset of links $C = \{ \cb_1,
\dots, \cb_n\}$, which we call point contacts. Each cut makes for one
network-incoming and one network-outgoing link where electric current
is  injected resp.\ drained by connecting the network to charge
reservoirs. The dynamics in the presence of the point contacts is
\cite{Janssen1999}
\begin{align}
  \left|\psi(t+1)\right> &= \UU \Big( Q \left|\psi(t) \right> +
    \sum\nolimits_{l=1}^n \left|\cb_l \right> a_l \Big) ,
\end{align}
where the projector $Q = 1 - \sum_{l=1}^n | \cb_l \rangle \langle
\cb_l|$ implements the draining action at the outgoing open ends, and
$a_l$ is the amplitude of the flux per time step fed into the incoming
end at $\cb_l$. We then consider stationary states of this
open-network dynamics. Without loss we take the quasi-energy to be zero, as
the statistical properties of the network model are
independent of it. We refer to the solutions of the stationarity
condition $\vert \psi(t+1) \rangle \equiv \vert \psi(t) \rangle$ as
\emph{scattering states}. For a system with $n$ point contacts, a
basis of scattering states is furnished by
\begin{align}
  \label{eq:scatt_state}
  \vert \psi_k \rangle \equiv \UU (1 - Q \UU )^{-1} \vert \cb_k \rangle
  \quad (k = 1, \ldots n) .
\end{align}
Note that $|\!| Q \UU |\!| < 1$, which ensures that the inverse exists
as a convergent power series $(1 - Q \UU)^{-1} = \sum_{t = 0}^\infty
(Q \UU)^t$. 

\paragraph{Main results and numerics.} The first result to be announced
is a statement about two-point functions, allowing one to measure the
scaling dimensions of primary fields. Consider a set of links
$R = \{ \rb_1,\dots, \rb_n\}$ for the purpose of (non-invasive)
observation, and define for $i, j, m =
1,\dots,n$:
\begin{align}\label{eq:An}
  A_m &= \Det K^{(m)} , \quad K_{ij}= \sum_{k=1}^n \psi_k(\rb_i)\,
  \overline{\psi_k(\rb_j)} ,
\end{align}
where $K^{(m)}$ denotes the upper-left $m\times m$ sub-matrix of $K$.
These observables are the open-network counterparts of those
considered in \cite{Gruzberg2013}. Suppose now that coarse graining of
the lattice takes the contact and observation regions ($C$ and $R$) to
single points, i.e.\ $\rb_i \to \rb$ and $\cb_i \to \cb$ for all $i$,
while $\rb$ and $\cb$ remain distinct. Denoting disorder averages by
$\EE \{ \ldots \}$ and CFT correlators as $\langle \ldots
\rangle$, we then claim that \cite{footnote1}
\begin{equation}\label{eq:2pt}
    \begin{split}
      &\EE\left\{ \left(A_1^{q_1-q_2} A_2^{q_2-q_3} \cdots
        A_n^{q_n}\right)(R,C) \right\} \\ &= a^{2\Delta_{q_1 \ldots\, q_n}}
    \big\langle \varphi_{q_1 \ldots\, q_n}(\rb) \, \Phi(\cb)
    \big\rangle \,,
    \end{split}
\end{equation}
where $q_1, \ldots, q_n$ are complex numbers, $\varphi_{q_1\dots\,
  q_n}$ is a CFT primary field with scaling dimension
$\Delta_{q_1\dots\, q_n}$, the operator $\Phi(\cb)$ represents the
contacts, and $a$ is the non-universal scale parameter of the
network. Even though $\Phi(\cb)$ is not a pure scaling field, it here
contributes a definite scaling dimension $\Delta_{q_1\dots\, q_n}$ due
to the orthogonality principle for two-point functions. Thus for an
infinite planar network we predict that the observable in
(\ref{eq:2pt}) depends on the distance between the contact and
observation regions as a pure power $\vert \rb - \cb
\vert^{-2\Delta_{q_1 \dots\, q_n}}$. For the special choice of $q_2 =
\dots = q_n = 0$ this prediction reduces to $\EE\left\{ |\psi_{\cb}
  (\rb)|^{2q_1} \right\} \propto \vert \rb - \cb \vert^{-2\Delta_{q_1
    0\, \dots\, 0}}$, which strongly suggests that $\Delta_{q_1 0\,
  \dots \, 0}$ coincides with the multi-fractality spectrum of the
local density of states \cite{Evers2008}. The analytical arguments
leading to (\ref{eq:2pt}) are sketched below.

Next we support our proposal by computing numerically some of the
observables above. We consider cylindrical networks of length $L =
400$ (with reflecting boundary conditions) and eight different
circumferences $W\in\{ 19, 22, \ldots, 40\}$.  We use
Eq.~\eqref{eq:scatt_state} to compute the scattering states for an
ensemble of $10^6$ disorder realizations. To illustrate the result
(\ref{eq:2pt}), we focus on the example of $\EE \{ A_n^q (R,C) \}$ for
$n$ contact and $n$ observation links. Assuming (\ref{eq:2pt}) and
using the CFT prediction for the correlator of (spinless) primary
fields on an infinite cylinder of width $W$ (see, e.g.,
\cite{DiFrancesco1997}), we have:
\begin{align}\label{eq:An-scaling}
  &\EE \{ A_n^q(R,C) \} = \alpha_{q,n}\,
  \zeta_{\rb\cb}^{-2\Delta_{q,n}} ,
  \\ \label{eq:zeta}
  &\zeta_{\rb\rb'} = \left| \frac{W}{\pi} \sinh \frac{\pi}{W} (\tau -
    \tau^\prime + \mathrm{i}\sigma - \mathrm{i}\sigma^\prime) \right| ,
\end{align}
where $\Delta_{q,n} \equiv \Delta_{q \dots q}$, and the form factor
$\alpha_{q,n}$ is due to the contact operator $\Phi(\cb)$. The
variables $\sigma$ and $\tau$ are the angular and longitudinal
cylindrical coordinates of $\rb$.
%

\begin{figure}
    \centering
    \includegraphics[width=8cm]{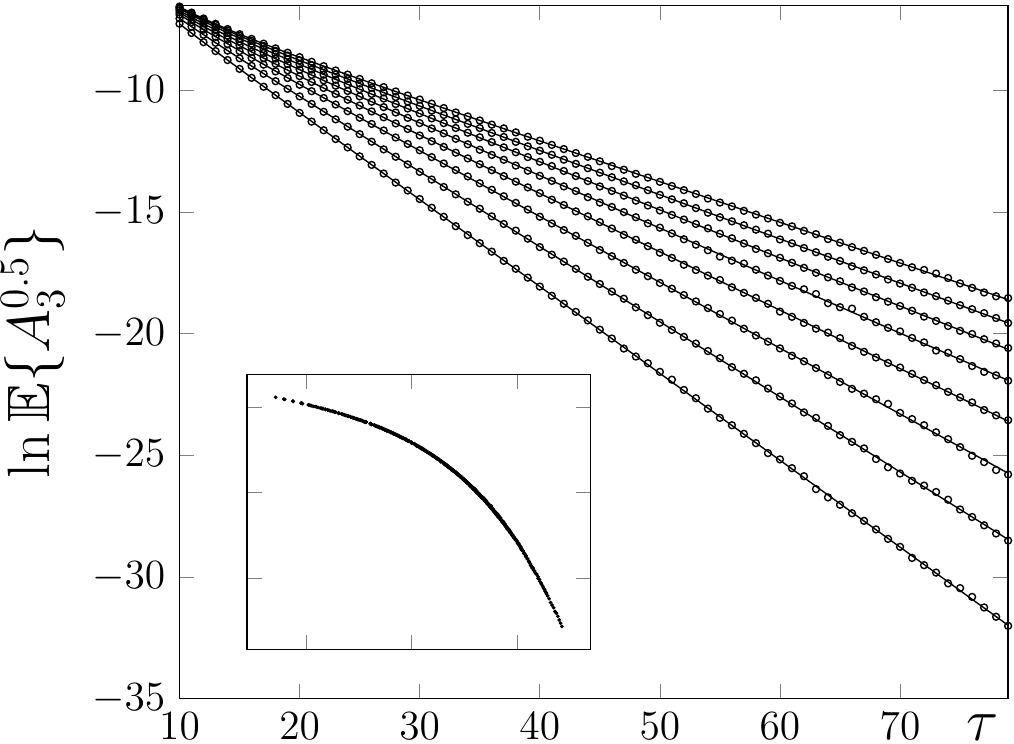}
    \caption{Fits of $\EE \{A_3^{0.5}\}$ by the CFT prediction
      \eqref{eq:An-scaling} for $W \in \{ 19, 22, \ldots, 40\}$ (bottom to
      top) as a function of $\tau = |\rb - \cb|$.  Inset: Collapse of
      rescaled curves $\ln \EE\{A_3^{0.5}\}+2\Delta_{0.5,3} \ln W$
      vs. $\ln x/W$ using $\Delta_{0.5,3} = 2.15$.} 
    \label{fig:A3}
\end{figure}

Detailed numerical investigations were performed of the correlator
$\EE \{ A_n^q (R,C) \}$ for $n = 1, 2, 3$. For $n > 1$ we place the
contacts on equivalent links of $n$ plaquettes next to each other; we
checked for $n = 2$ that our results do not change significantly when
this choice is modified. Data for the numerically most demanding case
of $n = 3$ are shown in Fig.\ \ref{fig:A3} for $q = 0.5$ as an
example. We see an excellent agreement over the whole range of
distances $|\rb - \cb|$ between the numerical data (circles) and the
functional behavior (solid line) predicted by \eqref{eq:An-scaling}.

In order to extract exponents and make an estimate of the statistical
errors, we use the following procedure. Given $q$ and $n$, we fit the
data for $\EE \{ A_n^q (R,C) \}$ by the prediction
\eqref{eq:An-scaling} for different $W$. This fit yields eight ``raw''
exponents $\Delta_{q,n}(W)$ from which we calculate the mean value and
the standard deviation. For $n = 3$ and $q = 0.5$, the inset of Fig.\
\ref{fig:A3} shows the data collapse by the optimal value of
$\Delta_{0.5,3}$ thus obtained. We interpret the excellent quality of
these fits as strong evidence that the $A_n^q$ indeed give rise to CFT
primary fields in the continuum limit.

\begin{figure}
    \centering
    \includegraphics[width=8cm]{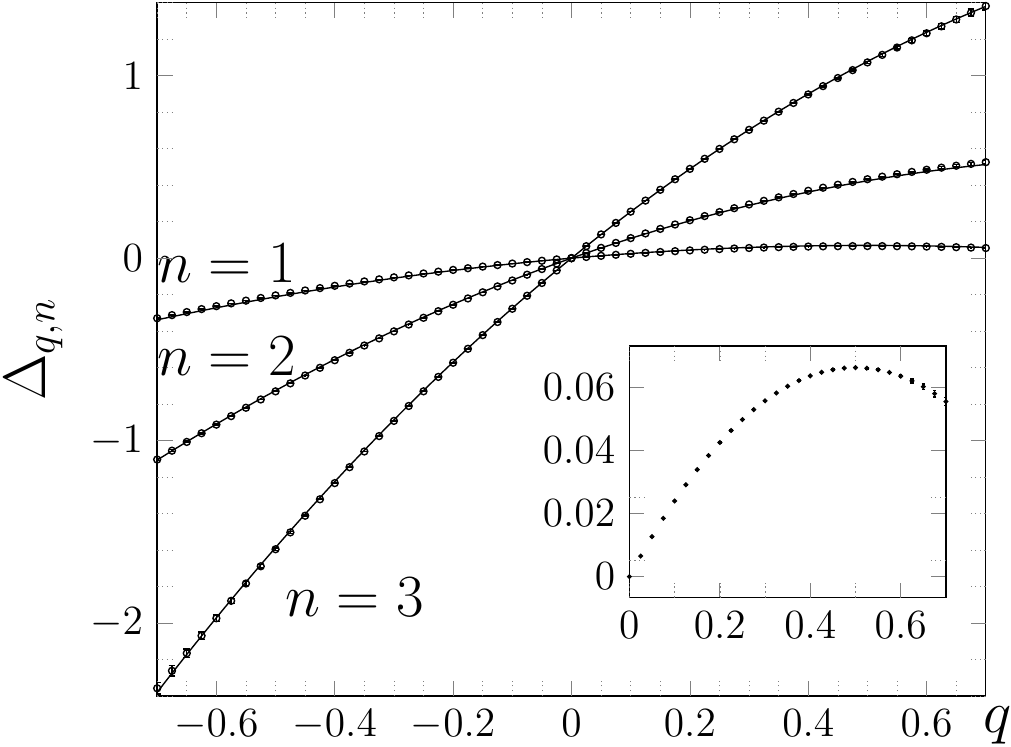}
 \caption{Scaling exponents
    $\Delta_{q,n}$ for $\EE \{ A_n^q \}$. The solid curves are plots
    of $\Delta_{q,n}=0.28 \, C_2 - 0.0011\, C_2^2 + 0.0014\, C_4$.
    Inset: zoom in for $n=1$.}
 \label{fig:alln}
\end{figure}

Having established the existence of these fields, we now turn to a
systematic analysis of the scaling dimensions $\Delta_{q,n}$. These
are constrained by $\Delta_{q,n} = \Delta_{n-q,n}$ due to an argument
\cite{Gruzberg2013} using Weyl group invariance. The simplest ansatz
compatible with that constraint would be $\Delta_{q,n} \propto
C_2(q,n)$, with $C_2 = n q (n - q)$ the quadratic Casimir eigenvalue
of the symmetry group at hand, but it is known from
\cite{Evers2008b,Obuse2008} that this so-called ``parabolic
approximation'' needs improvement by including in $\Delta_{q,1}$ the
square of $C_2$ with a small coefficient. Focusing on
$n = 1$, we find that $\Delta_{q,1}$ is in fact described reasonably
well by the parameter set of \cite{Evers2008b,Obuse2008}. The
situation changes, however, when we take into account our results for
$n = 2, 3$, as shown in Fig.\ \ref{fig:alln}.

To get a good fit of all data $n = 1, 2, 3$ simultaneously, we find it
necessary to include in $\Delta_{q,n}$ the Casimir eigenvalue of
degree four \cite{Scheunert}. Evaluated on $A_n^q$ this is
\begin{equation}
    C_4(q,n) = - n (q (n - q))^2 + n (n^2 - 1/2) q (n-q) .
\end{equation}
We leave it for future work to decide whether this is a real effect or
might have another explanation, e.g.\ by the presence of irrelevant
operators perturbing the CC model away from the CFT fixed point
\cite{Obuse2013}.

\begin{figure}
    \centering
    \includegraphics[width=8cm]{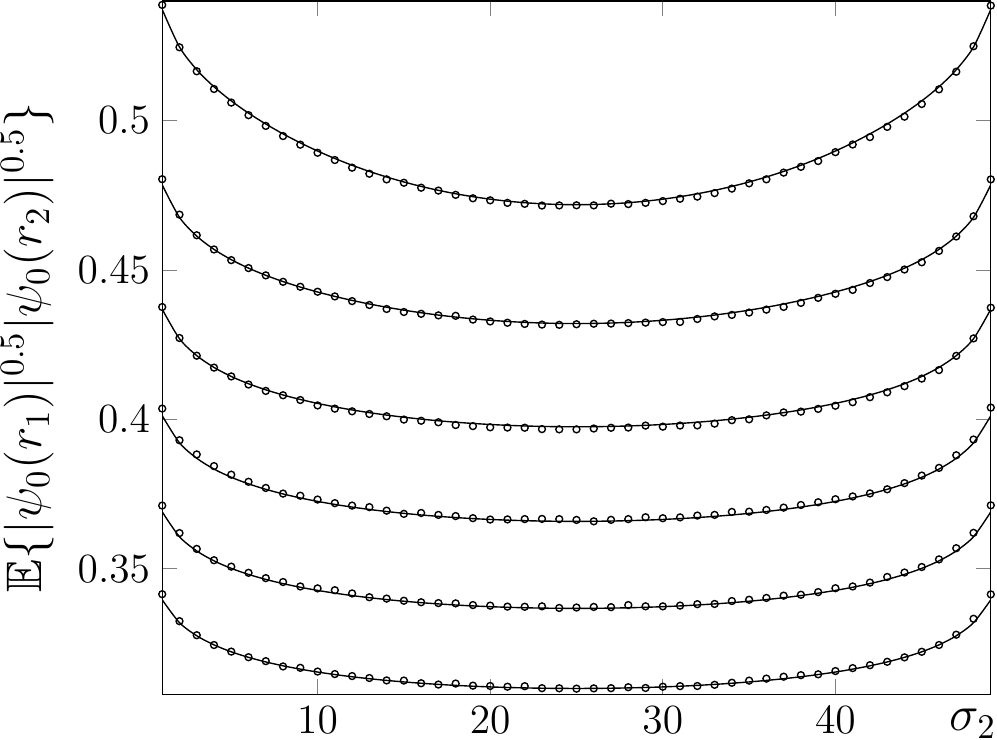}
    \caption{Comparison between the
    CFT prediction \eqref{eq:3pt-scaling} and the angular dependence
    of the 3-point function \eqref{eq:3pt-scalar} computed for $W =
    50$ and $\tau_1 = \tau_2 = 10, 20, \ldots, 60$ (top to bottom).}
  \label{fig:3pt}
\end{figure}

To strengthen our claim that the operators in (\ref{eq:2pt}) behave as
CFT primary fields, we present a second result, this time for a
three-point function. Here we sacrifice generality for simplicity
and open the network at just a single contact link $\cb_0$ to study
the scalar-type observable $A_1(\rb) = \vert \psi_0 (\rb) \vert^2$ at
two observation links $\rb_1$ and $\rb_2$. Our assertion is that,
after coarse graining,
\begin{equation}\label{eq:3pt-scalar}
  \EE\left\{ |\psi_0(\rb_1)|^{2q_1} |\psi_0(\rb_2)|^{2q_2} \right\}
  \propto \left\langle \varphi_{q_1}(\rb_1) \varphi_{q_2}(\rb_2)
    \Phi(\cb_0) \right\rangle
\end{equation}
depends on $\rb_1, \rb_2, \cb_0$ as a CFT three-point function.
Because of the special nature of the operators $\varphi_{q_1}$,
$\varphi_{q_2}$ as ``highest-weight vectors'' (see below), the contact
operator $\Phi(\cb_0)$ still contributes a definite scaling dimension
$\Delta_{q_0}$, where $q_0=q_1 + q_2$. The CFT prediction for
three-point functions \cite{DiFrancesco1997} then gives
\begin{equation}
  \label{eq:3pt-scaling}
  \begin{split}
    &\EE\left\{|\psi_0(\rb_1)|^{2 q_1} |\psi_0(\rb_2)|^{2 q_2} \right\}
    \propto \\
    & \times\zeta_{\rb_1\rb_2}^{-\Delta_{q_1}-\Delta_{q_2}+\Delta_{q_0}}
    \zeta_{\rb_2\cb_0}^{-\Delta_{q_2}-\Delta_{q_0}+\Delta_{q_1}}
    \zeta_{\cb_0 \rb_1}^{-\Delta_{q_0}-\Delta_{q_1}+\Delta_{q_2}}\, ,
  \end{split}
\end{equation}
where $\zeta_{\rb\rb'}$ was defined in Eq.~\eqref{eq:zeta}. In our
numerical test we take $\rb_1$ and $\rb_2$ to have the same
$\tau$-coordinates and $\rb_1$ to share the $\sigma$-coordinate of the
point contact $\cb_0$, while $\rb_2$ moves along the circumference of
the cylinder. Fig.\ \ref{fig:3pt} shows the angular dependence
observed for $W = 50$ together with the prediction
\eqref{eq:3pt-scaling} at $q_1 = q_2 = 1/4$. The exponents
$\Delta_{1/4,1}$ and $\Delta_{1/2,1}$ used are those extracted from
the study of the two-point function.

\paragraph{Analytical argument.} Finally, we sketch the reasoning that
leads to Eqs.~(\ref{eq:2pt}), (\ref{eq:3pt-scalar}); details and
generalizations will be discussed in \cite{Longpaper}. We use a
variant of the Efetov-Wegner method to pass from the CC model to a
supersymmetric vertex model (see e.g.\
\cite{Gruzberg1997,Zirnbauer1997}) as follows. For each link $\rb$ of
the network we introduce $n$ replicas of charged $(\pm)$ canonical
bosons and fermions: $b_{\pm,k} (\rb)$, $f_{\pm,k} (\rb)$ ($k = 1,
\ldots, n$), acting on a Fock space with vacuum $\vert 0_{\rb}
\rangle$. The time-evolution operator $U$ of the closed network is
then replaced by its second quantization $\rho(U)$ acting on the
tensor product of all these Fock spaces. This factors as $\rho(U
\equiv \mathrm{e}^X) = \rho_+( \mathrm{e}^X) \rho_-(\mathrm{e}^X)$
where, assuming the summation convention,
\begin{equation}
  \begin{split}
    \rho_+ (\mathrm{e}^X) &= \mathrm{e}^{b^\dagger_{+,k} (\rb)
      X_{\rb\rb'} b_{+,k} (\rb') + f^\dagger_{+,k} (\rb) X_{\rb\rb'}
      f_{+,k} (\rb')}, \\ \rho_- (\mathrm{e}^X) &= \mathrm{e}^{-
      b_{-,k} (\rb) X_{\rb\rb'} b_{-,k}^\dagger(\rb') + f_{-,k}(\rb)
      X_{\rb\rb'} f_{-,k}^\dagger (\rb')} .
  \end{split}
\end{equation}
It is important that second quantization preserves operator
products. In particular, $\rho(U_{\rm s} U_{\rm r}) = \rho(U_{\rm s})
\rho(U_{\rm r})$. Statistical averages in this Fock representation are
defined by $\langle A \rangle_\mathcal{F} := \STr \rho(U) A$, where
$\STr$ is the supertrace over the total Fock space.

For simplicity, we now specialize to $n = 1$ and return to $n \geq 1$
below. Given $Q_\varepsilon = Q + \varepsilon \vert \cb \rangle
\langle \cb \vert$, let
\begin{equation}
    \pi_0(\cb) = \vert 0_{\cb} \rangle \langle 0_{\cb} \vert =
    \lim_{\varepsilon \to 0+} \rho_+(Q_\varepsilon) \rho_-(Q_\varepsilon^{-1})
\end{equation}
be the projector on the vacuum state at $\cb$. The second-quantized
formalism is connected to observables of the first-quantized network
model by the basic identities
\begin{equation}\label{eq:BasicID}
  \begin{split}
    \langle b_+^\dagger (\rb) b_+(\rb) \, \pi_0(\cb) \rangle_\mathcal{F}
    &= \langle \rb \vert Q U (1 - Q U)^{-1} \vert \rb \rangle , \\
    \langle b_-(\rb) b_-^\dagger(\rb) \, \pi_0(\cb) \rangle_\mathcal{F}
    &= \langle \rb \vert (1 - U^{-1} Q)^{-1} \vert \rb \rangle .
  \end{split}
\end{equation}
To exploit these, we introduce the following key objects for an
observation link $\rb$:
\begin{align}
    &\ZZ_q(\rb,\cb) = \langle (B^\dagger B)^q (\rb) \, \pi_0(\cb)
    \rangle_\mathcal{F} , \label{eq:13} \\ &B^\dagger = b_+^\dagger -
    \mathrm{e}^{\mathrm{i} \alpha} b_- \,, \quad B = b_+ -
    \mathrm{e}^{-\mathrm{i} \alpha} b_-^\dagger \,,
\end{align}
where $\mathrm{e}^{\mathrm{i} \alpha}$ is any (fixed) unitary
number. Note the vanishing commutator $[ B , B^\dagger ] = 0$. Note
also that neither $B$ nor $B^\dagger$ annihilates any state in Fock
space. Therefore the operator $B^\dagger B$ is strictly positive and
$(B^\dagger B)^q$ makes good sense for any $q \in \mathbb{C}$.
Moreover, Wick's theorem holds in our noninteracting-particle
situation before disorder averaging. Thus for $q$ a positive integer
we have
\begin{equation}\label{eq:15}
    \ZZ_q(\rb,\cb) = q! \, \ZZ_1(\rb,\cb)^q .
\end{equation}
This extends to complex $q$ by analytic continuation. The basic
correlator $\ZZ_1(\rb,\cb)$ is expressed in terms of the scattering
state $\vert \psi_{\cb} \rangle$ by the following computation; it is
based on the identities (\ref{eq:BasicID}) and in the last step uses
that scattering states with incoming-wave and outgoing-wave
boundary conditions are unitarily related to each other:
\begin{align}
    &\ZZ_1(\rb,\cb) = \langle (b_+^\dagger b_+^{\vphantom{\dagger}} +
    b_-^{\vphantom{\dagger}} b_-^\dagger)(\rb) \, \pi_0(\cb) \rangle_\mathcal{F} \cr
    &= \langle \rb \vert (1 - U^{-1} Q)^{-1} U^{-1} (1 - Q ) U (1 - Q
    U)^{-1} \vert \rb \rangle \cr &= \left\vert \langle \rb \vert
      (1 - U^{-1} Q)^{-1} U^{-1} \vert \cb \rangle \right\vert^2 =
    \vert \psi_{\cb}(\rb) \vert^2 \label{eq:16} .
\end{align}

Next we take the disorder average. This is straightforward in the Fock
representation since $\rho(U)=\rho(U_\mathrm{s}) \rho (U_\mathrm{r})$
and averaging over the random phases in $\rho (U_\mathrm{r})$ simply
kills all states with non-zero charge. For any charge-conserving
operator $A$ we thus obtain
\begin{equation}
    \EE \left\{ \langle A \rangle_\mathcal{F} \right\} =
    \EE \left\{ \STr A \rho(U_\mathrm{s}) \rho(U_\mathrm{r}) \right\}
    = \STr^\prime A \rho(U_\mathrm{s}) ,
\end{equation}
where $\STr^\prime$ is $\STr$ restricted to the zero-charge sector. In
this way we arrive at what is called a \emph{vertex model}. We denote
vertex-model averages by $\langle A \rangle_\mathcal{V} \equiv
\STr^\prime A \rho(U_\mathrm{s})$.

The operators $B^\dagger B$ and $\pi_0$ conserve charge, so by taking
the disorder average of \eqref{eq:13} and using
(\ref{eq:15},\ref{eq:16}) we get
\begin{equation}
    \EE \left\{ \vert \psi_{\cb}(\rb) \vert^{2q} \right\} = q!^{-1}
    \langle (B^\dagger B)^q (\rb) \, \pi_0(\cb) \rangle_\mathcal{V} .
\end{equation}
This is an exact result. Although our focus has been on $n = 1$, the
general case $n \geq 1$ can be handled in a similar way. The outcome
is an exact relation expressing network-model averages as vertex-model
averages:
\begin{align}\label{eq:19}
\begin{split}
  &\EE \left\{ ( A_1^{q_1-q_2} A_2^{q_2-q_3} \cdots A_n^{q_n}) (R,C)
  \right\} \\
  &= f(q) \langle (D_1^{q_1-q_2} D_2^{q_2-q_3} \cdots
  D_n^{q_n}) (R) \, \pi_0 (C) \rangle_\mathcal{V} ,
\end{split}\\
  &D_m(R) =
  \Det \left( \sum\nolimits_{k=1}^m B_k^\dagger(\rb_i) B_k(\rb_j)
  \right)_{i,j=1,\ldots,m} .
\end{align}
Here $f(q)$ is a combinatorial factor. The determinants $D_m(R)$ are
well-defined because the matrix elements $\sum B_k^\dagger(\rb_i)
B_k(\rb_j)$ all commute. The analytic continuation to complex powers
of $D_m$ is well-defined because $D_m > 0$.

We will now argue that the expression in \eqref{eq:19} becomes a pure
scaling function in the continuum limit. The key ingredient here is
symmetry: the statistical average $\langle \ldots \rangle_\mathcal{V}$
is invariant under the global action of a group with Lie superalgebra
$\mathfrak{g} \equiv \mathfrak{gl} (2n|2n)$ generated by all
charge-conserving bilinears in $b_{\pm,k}$, $f_{\pm,k}$, and their
adjoints.

What is most remarkable about the operators $D_m$ is their property of
being \emph{highest-weight vectors} for $\mathfrak{g}$. By this we
mean that there exists a maximal abelian subalgebra $\mathfrak{h}
\subset \mathfrak{g}$ such that the $D_m$ are (i) eigenoperators
w.r.t.\ the commutator action by all generators from $\mathfrak{h}$
and (ii) are annihilated by all the raising operators, i.e.\ the
operators from $\mathfrak{g}$ which are positive root vectors for
$\mathfrak{h}$. Since the operation of taking the commutator satisfies
the Leibniz rule, these properties carry over to products of powers of
$D_m$. Thus $\varphi_{q_1 \ldots q_n}^\mathrm{lat}(R) \equiv D_1^{q_1
  - q_2}(R) D_2^{q_2 - q_3}(R) \cdots D_n^{q_n}(R)$ is a
highest-weight vector for $\mathfrak{g}$. The operators $\varphi_{q_1
  \ldots q_n}^\mathrm{lat}(R)$ for different $q = (q_1, \ldots, q_n)$
(modulo Weyl transformations) lie in inequivalent representations of
$\mathfrak{g}$. Therefore, by a Schur lemma argument they cannot be
mixed by the transfer matrix of the $\mathfrak{g}$-invariant vertex
model. Thus we expect them to become pure scaling fields of the
renormalization flow in the continuum limit. (This has to taken with a
grain of salt since our CFT has a logarithmic sector
\cite{Saleur2007}.)

Finally, our methods generalize to any multi-point function of the
scaling operators given here. In particular, one can derive
Eq.~\eqref{eq:3pt-scalar} by the reasoning above. To arrive at
\eqref{eq:3pt-scaling} with $\Delta_{q_0} = \Delta_{q_1 + q_2}$ one
uses that the product of two highest-weight vectors with weights $q_1$
and $q_2$ is another highest-weight vector with weight $q_0 = q_1 +
q_2$.

\paragraph{Summary and outlook.} We have presented new methods and
relations by which to make a systematic study of the IQH plateau
transition. For the first time in the context of Anderson transitions,
we have identified and studied operators whose correlation functions
decay as pure powers at criticality and thus are candidates for CFT
primary fields. Although we applied our techniques to the specific
case of the CC model, they are rather general and can be used for
other situations as well.  Future applications will include the spin
quantum Hall transition \cite{Gruzberg1999} and the study of boundary
criticality \cite{Bondesan2012}.

\paragraph{Acknowledgment.} We thank R.\ Klesse and A.W.W.\ Ludwig for
useful discussions and the former also for help with the numerical
simulations. Financial support by DFG grant ZI-513/1-2 is gratefully
acknowledged.


\begin{thebibliography}{40}
%
\bibitem{Evers2008} F. Evers, A.D. Mirlin, Rev. Mod. Phys. {\bf 80},
  1355 (2008).
%
\bibitem{Bhaseen2000} M.J. Bhaseen, I.I. Kogan, O.A. Soloviev,
  N. Taniguchi, A.M. Tsvelik, Nucl. Phys. B {\bf 580}, 688 (2000).
%
\bibitem{Ikhlef2011} Y. Ikhlef, P. Fendley, J. Cardy, Phys. Rev. B
  {\bf 84}, 144201 (2011).
%
\bibitem{Bettelheim2012} E. Bettelheim, I.A. Gruzberg, A.W.W. Ludwig,
  Phys. Rev. B {\bf 86}, 165324 (2012).
%
\bibitem{Janssen1999} M. Janssen, M. Metzler, M.R. Zirnbauer,
  Phys. Rev. B {\bf 59}, 15836 (1999).
%
\bibitem{Klesse2001} R. Klesse, M.R. Zirnbauer, Phys. Rev. Lett. {\bf
    86}, 2094 (2001).
%
\bibitem{Obuse2013} H. Obuse, S. Bera, A.W.W. Ludwig, I.A. Gruzberg,
  F. Evers, Europhys. Lett. {\bf 104}, 27014 (2013).
%
\bibitem{Gruzberg2011} I.A. Gruzberg, A.W.W. Ludwig, A.D. Mirlin,
  M.R. Zirnbauer, Phys. Rev. Lett. {\bf 107}, 086403 (2011).
%
\bibitem{Gruzberg2013} I.A. Gruzberg, A.D. Mirlin, M.R. Zirnbauer,
  Phys. Rev. B {\bf 87}, 125144 (2013).
%
\bibitem{Zirnbauer1997} M.R. Zirnbauer, J. Math. Phys. {\bf 38}, 2007
  (1997).
%
\bibitem{Gruzberg1997} I.A. Gruzberg, N. Read, S. Sachdev,
  Phys. Rev. B {\bf 55}, 10593 (1997).
%
\bibitem{Chalker1988} J.T. Chalker, P.D. Coddington, J. Phys. C: Solid
  State Physics {\bf 21}, 2665 (1988).
%
\bibitem{footnote1} Eqs.~(\ref{eq:2pt},\ref{eq:3pt-scalar}) are meant
  to hold on large enough scales, where the contribution of irrelevant
  operators \cite{Obuse2013} is negligible.
 %
\bibitem{DiFrancesco1997} P. Di Francesco, P. Mathieu, D. Senechal,
  \textit{Conformal Field Theory} (Springer, Berlin, 1997).
%
\bibitem{Evers2008b} F. Evers, A. Mildenberger, A.D. Mirlin,
  Phys. Rev. Lett. {\bf 101}, 116803 (2008).
%
\bibitem{Obuse2008} H. Obuse, A.R. Subramaniam, A. Furusaki,
  I.A. Gruz\-berg, A.W.W. Ludwig, Phys. Rev. Lett. {\bf 101}, 116802
  (2008).
%
\bibitem{Scheunert} M.\ Scheunert, J. Math. Phys. {\bf 24}, 2681 (1983).
%
\bibitem{Longpaper} R.\ Bondesan, D.\ Wieczorek, and M.R.\ Zirnbauer (in
  preparation).
%
\bibitem{Saleur2007} N. Read, H. Saleur, Nucl. Phys. B {\bf 777}, 316
  (2007).
%
\bibitem{Gruzberg1999} I.A. Gruzberg, A.W.W. Ludwig, N. Read,
  Phys. Rev. Lett. {\bf 82}, 4524 (1999).
%
\bibitem{Bondesan2012} R. Bondesan, I.A. Gruzberg, J.L. Jacobsen,
  H. Obuse, H. Saleur, Phys. Rev. Lett. {\bf 108}, 126801 (2012).
\end{thebibliography}
\end{document}